%% file: paper.tex
\documentclass[journal,twoside]{IEEEtran}
\pdfoutput=1
\usepackage[utf8x]{inputenc}
\usepackage{amsfonts, amsmath, amssymb, array}
\usepackage{graphicx}
\usepackage{pgfplots}
\usepackage{tikz}
\pgfplotsset{compat=newest}
\pgfplotsset{plot coordinates/math parser=false}

\usepackage{nohyperref}

\usepackage{color}
\usepackage{import} 
\usepackage{bm} 
\usepackage{bigints} 
\usepackage{amssymb, amsfonts, mathtools}
\usepackage{cases}
\usepackage{cite}

\usepackage{amsthm}

\usepackage[noabbrev]{cleveref} 
\crefformat{equation}{(#1)} 
\Crefformat{equation}{(#1)} 
\let\origref\ref 
\let\ref\Cref    

\newlength\figureheight
\newlength\figurewidth

\input{math_header}
\input{defines}

\begin{document}

\title{Quickest Eigenvalue-Based Spectrum Sensing using Random Matrix Theory}

\author{\IEEEauthorblockN{Martijn~Arts,~%
        Andreas~Bollig~%
        and~Rudolf~Mathar\\}%
        \IEEEauthorblockA{Institute for Theoretical Information Technology,~RWTH Aachen University,~D-52074 Aachen,~Germany\\
        E-mail: \{arts, bollig, mathar\}@ti.rwth-aachen.de}
        \thanks{This work was partly supported by the Deutsche Forschungsgemeinschaft (DFG) projects CoCoSa (grant MA 1184/26-1).}
        \thanks{This work has been submitted to the IEEE for possible publication. Copyright may be transferred without notice, after which this version may no longer be accessible.}
        \thanks{\textcopyright~2014, 2015 IEEE. Personal use of this material is permitted. Permission from IEEE must be obtained for all other uses, including reprinting/republishing this material for advertising or promotional purposes, collecting new collected works for resale or redistribution to servers or lists, or reuse of any copyrighted component of this work in other works.}}

\maketitle

\input{content}

\bibliographystyle{ieeetr}
\bibliography{literature.bib}

\end{document}

%% file: math_header.tex
\newcommand{\fac}[1]{\ensuremath{#1!}}

\newcommand{\abs}[1]{\ensuremath{\left\vert #1 \right\vert}}
\newcommand{\conj}[1]{\ensuremath{#1^{*}}}

\newcommand{\ud}{\ensuremath{\,\mathrm{d}}} 
\newcommand{\ppart}[1]{\ensuremath{\left[#1\right]^{+}}}

\newcommand{\N}{\ensuremath{\mathbb{N}}}
\newcommand{\C}{\ensuremath{\mathbb{C}}}  

\renewcommand{\vec}[1]{\ensuremath{\bm{\MakeLowercase{#1}}}}
\newcommand{\mat}[1]{\ensuremath{\bm{\MakeUppercase{#1}}}}

\newcommand{\matc}[2]{\ensuremath{\bm{\MakeLowercase{#1}}_{#2}}} 
\newcommand{\transp}[1]{\ensuremath{#1^{\scriptscriptstyle\bm T}}}
\newcommand{\herm}[1]{\ensuremath{#1^{\scriptscriptstyle\bm H}}}

\newcommand{\I}{\mat{i}}

\newcommand{\zeromat}{\mat{0}}

\newcommand{\onevec}{\vec{1}}

\newcommand{\hadamard}{\otimes}

\newcommand{\bessel}[1]{\ensuremath{\mathcal{I}_{#1}}}
\newcommand{\hypgeozero}[3]{\ensuremath{{_{0}\mathcal{F}_{#1}}(#2;#3)}}

\newcommand{\pochhammer}[2]{\ensuremath{\left(#1\right)_{#2}}}

\newcommand{\dist}{\ensuremath{\sim}}
\newcommand{\normal}[2]{\ensuremath{\mathcal{N}(#1, #2)}}
\newcommand{\SCN}[2]{\ensuremath{\mathcal{CN}(#1, #2)}}

\newcommand{\CWUN}[3]{\ensuremath{\mathcal{CW}_{#1}(#2, #3)}}
\newcommand{\CWUC}[2]{\ensuremath{\mathcal{CW}_{#1}(#2)}}

\newcommand{\expop}{\ensuremath{\mathrm{E}}}

\newenvironment{bigequations}{\begin{figure*}[!ht] \normalsize}{\end{figure*}}

%% file: defines.tex
\newcommand{\dB}{\text{dB}}

\newcommand{\dummypar}{a}
\newcommand{\dummyparB}{b}

\newcommand{\cov}{r}
\newcommand{\sample}{y}
\newcommand{\signal}{x}
\newcommand{\noise}{w}
\newcommand{\sigsymbol}{s}
\newcommand{\ncm}{\Omega}
\newcommand{\snr}{\alpha}
\newcommand{\numusers}{K}
\newcommand{\numsamples}{N}
\newcommand{\hypzero}{\mathcal{H}_0}
\newcommand{\hypone}{\mathcal{H}_1}
\newcommand{\timeidx}{t}
\newcommand{\blockidx}{k}

\newcommand{\teststat}{T}
\newcommand{\threshold}{h}
\newcommand{\eig}{\lambda}
\newcommand{\eigncm}{\omega}

\newcommand{\pd}{P_\text{d}}
\newcommand{\pfa}{P_\text{fa}}

\newcommand{\tchange}{t_\text{c}}
\newcommand{\tdetect}{t_\text{d}}
\newcommand{\td}{\tau_\text{d}}
\newcommand{\tfa}{\tau_\text{fa}}

\newcommand{\llr}{l}
\newcommand{\cusum}{g}

\newcommand{\cusumseq}{Z}
\newcommand{\cusumsample}{z}
\newcommand{\cusumpdfvar}{\cusumsample}
\newcommand{\cusumpdfpar}{\theta}

\newcommand{\cusumsup}{\gamma}
\newcommand{\cusumsupshift}{\delta}
\newcommand{\cusumroot}{\varphi}

\newcommand{\glrt}{l_G}
\newcommand{\glr}{g_G}
\newcommand{\glrvar}{m}
\newcommand{\glrsnr}{\protect{\hat{\snr}}}

\newcommand{\idxvar}{j}
\newcommand{\idxvarB}{l}

\newcommand{\sumapprox}{\MakeUppercase{\idxvar}_s}

\newcommand{\pdf}{f}
\newcommand{\pdfvar}{\teststat}
\newcommand{\pdfconst}{K}
\newcommand{\cdf}{F}
\newcommand{\cdfvar}{\pdfvar}

\newcommand{\matdummy}{A}

\newcommand{\matridx}{i}
\newcommand{\matcidx}{j}

\newcommand{\ucdummyf}{\Delta}
\newcommand{\ucdummyfpOne}{a}
\newcommand{\ucdummyfpTwo}{b}
\newcommand{\ucdummyfpThree}{c}

\newcommand{\ucdummycf}{p}

\newcommand{\undummyf}{\beta}

\newcommand{\pdfconstexp}[1]{\xi(#1)}

\newcommand{\vandmat}[1]{\mat{V_1}(#1)}
\newcommand{\hypgeomat}[2]{\mat{F}(#1,#2)}

\renewcommand{\det}[1]{\ensuremath{\left\vert #1 \right\vert}}

%% file: content.tex
\begin{abstract}
We investigate the potential of quickest detection based on the eigenvalues of the sample covariance matrix for spectrum sensing applications. A simple phase shift keying (PSK) model with additive white Gaussian noise (AWGN), with $1$ primary user (PU) and $\numusers$ secondary users (SUs) is considered. Under both detection hypotheses $\hypzero$ (noise only) and $\hypone$ (signal + noise) the eigenvalues of the sample covariance matrix follow Wishart distributions. For the case of $\numusers = 2$ SUs, we derive an analytical formulation of the probability density function (PDF) of the maximum-minimum eigenvalue (MME) detector under $\hypone$. Utilizing results from the literature under $\hypzero$, we investigate two detection schemes. First, we calculate the receiver operator characteristic (ROC) for MME block detector based on analytical results. Second, we introduce two eigenvalue-based quickest detection algorithms: a cumulative sum (CUSUM) algorithm, when the signal-to-noise ratio (SNR) of the PU signal is known and an algorithm using the generalized likelihood ratio, in case the SNR is unknown. Bounds on the mean time to false-alarm $\tfa$ and the mean time to detection $\td$ are given for the CUSUM algorithm. Numerical simulations illustrate the potential advantages of the quickest detection approach over the block detection scheme.
\end{abstract}

\section{Introduction}
\IEEEPARstart{M}{odern} telecommunication systems face ever-increasing demands on data rates in order to support a growing number of mobile applications, e.g., mobile audio and video streaming. Since unlicensed wireless frequency spectrum is already very rare, it has been proposed to reuse parts of the spectrum which are already licensed. This is due to the fact that large parts of the spectrum are not in use in certain geographical locations or at certain points in time. The term dynamic spectrum access groups efforts to make use of such spectrum holes in order to increase bandwidth, see \cite{zhao_survey_2007} for an overview. One subgroup of research focuses on opportunistic spectrum access, in which the unlicensed users, called secondary users (SUs), determine the presence of licensed primary users (PUs) and based on these findings decide whether
to start unlicensed communication in an autonomous fashion. Obviously, a major challenge is the reliable detection of transmission opportunities for the SUs, such that ideally no or only a negligible amount of interference is caused for PUs. 

Detecting spectrum holes is known as \emph{spectrum sensing} in the literature and there have been a number of publications on detection schemes based on various signal features, see \cite{axell_spectrum_2012} for a fairly recent review. Among those features, the sample covariance matrix and particularly its eigenvalues have been used in order to design several detectors, see e.g., \cite{zeng_maximum-minimum_2007,zeng_eigenvalue-based_2009, zeng_blindly_2008, yang_blind_2011, font-segura_quadratic_2013, bollig_mmme_2013}. A well known one, which will be relevant in this work, is the ratio of the maximum to minimum eigenvalue (MME), i.e. the standard condition number (SCN) of the sample covariance matrix, see \cite{zeng_maximum-minimum_2007, zeng_eigenvalue-based_2009}. Since the noise power of the receiver is contained in both eigenvalues, it is canceled out in the ratio. Thus, designing an alarm threshold for this detector is possible without knowledge of the noise power.

When analyzing communication systems, receiver noise is often modeled as additive white Gaussian noise (AWGN). If the entries of a matrix $\mat{\matdummy}$ are Gaussian distributed, then the matrix $\mat{\matdummy}\herm{\mat{\matdummy}}$ is called a Wishart matrix \cite{james_distributions_1964}. On the one hand, such matrices occur in the analysis of multiple-input multiple-output (MIMO) communication systems as the instantaneous MIMO correlation matrix. On the other hand, Wishart matrices are also encountered when using the sample covariance matrix for spectrum sensing detectors, as will be explained in greater detail in \ref{sec:system_model}. The field of \emph{Random Matrix Theory} (RMT) studies said Wishart matrices and in the recent years there has been significant progress on the exact distributions of their eigenvalues, see e.g., \cite{zanella_marginal_2009} for a summary. Based on the joint ordered eigenvalue distribution, also the distribution of the SCN can be found, see \cite{kortun_performance_2011, matthaiou_condition_2010}. This result, among other approximating approaches from RMT, has recently been applied to spectrum sensing \cite{zhang_spectrum_2012}. In this work, we will utilize the SCN distribution in a spectrum sensing problem as well.

The spectrum sensing detectors discussed above are most often employed with a fixed sample size. There, a block of samples is taken, the test statistic is calculated and subsequently a decision is made by comparing the value of the test statistic to a predetermined threshold. One disadvantage of said \emph{block detection} schemes lies in the fact that the number of samples taken is independent of the detection difficulty of the current situation. For example, even if a very strong PU signal is present, it will take as many samples until a decision is reached as originally planned when defining the threshold. This may introduce significant delays, which may either result in interference for the PUs or reduce the data rate by shortening the transmission window of the SUs. 

\begin{figure}
    \centering
    \def\svgwidth{0.7\columnwidth}
    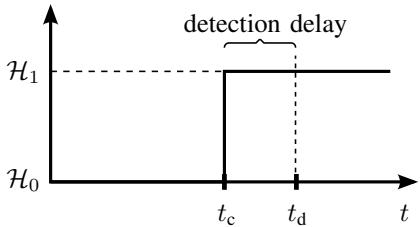
    \caption{Quickest detection problem. At the change time $\tchange$, hypothesis $\hypone$ becomes true. The detection algorithm gives an alarm at the detection time $\tdetect$, hence the detection delay or time to detection is $\tdetect - \tchange$.}
    \label{fig:QD}
\end{figure}

If one seeks to minimize the detection delay, the hypothesis detection problem can be cast within the framework of \emph{quickest detection}. In contrast to block detection, it is assumed that it is known which hypothesis is true and that an upcoming change is to be detected with minimum delay. Let $\hypzero$ identify the situation that the PU is not transmitting and only noise is received. Consequently, let $\hypone$ define the situation that the PU is transmitting and a signal with additive noise is received. Consider without loss of generality that there is no PU transmitting (hypothesis $\hypzero$ is true). At the unknown change time $\tchange$ a PU starts transmitting, such that the corresponding hypothesis $\hypone$ becomes true and at the detection time $\tdetect$ the detection algorithm raises an alarm, see also \ref{fig:QD}. 
The objective in a quickest detection problem is to minimize the mean detection delay $\td$, while having a mean time to false alarm $\tfa$ that is greater than a predefined value. Quickest detection has been applied to the spectrum sensing problem and is also known as \emph{quickest spectrum sensing} in the literature. The first publications in this field were \cite{lai_quickest_2008, li_quickest_2008} using Gaussian noise \& signal and Gaussian noise \& sinusoidal signal models, respectively. Quickest spectrum sensing using cyclostationary signal features was investigated in \cite{li_cyclostationary_2010}. Collaboration between multiple SUs using quickest detection has been studied in \cite{zarrin_cooperative_2009, li_collaborative_2010}. Also, the case of an SU having multiple antennas was analyzed in \cite{hanafi_extension_2013}.

To the best of our knowledge, so far there has been no investigation whether it is feasible to use quickest spectrum sensing, while using a test statistic that is based on the sample covariance matrix or its eigenvalues. In this work, we use a very basic model using two receivers that can either be interpreted as one SU with $\numusers$ receiving antennas or as $\numusers$ collaborating SUs in close proximity. That is, they are assumed to have the same signal-to-noise ratio (SNR). It is furthermore assumed, that only one PU is potentially present using phase shift keying (PSK) signal modulation. With this simple model, we study eigenvalue-based quickest spectrum sensing on the basis of the MME test statistic mentioned above. The main contributions and the outline of this paper can be summarized as follows. First, we formalize the signal model and specify the Wishart matrices involved under both hypotheses in \ref{sec:system_model}. Second, we develop the exact probability distribution function (PDFs) of the MME test statistic under hypothesis $\hypone$ for the special case of 2 SUs using results from RMT in \ref{sec:mme_distributions}. For the hypothesis $\hypzero$ we recall results known from literature. We also give details on the numerical evaluation of the PDFs in the same Section. Then, we apply these results to predict the performance of the MME block detection algorithm by theoretical calculation of the receiver operator characteristic (ROC) in \ref{sec:BD}. In \ref{sec:QD} we study the quickest detection problem using the MME test statistic with and without knowledge of the SNR. For the first case theoretical performance bounds are given. A numerical performance evaluation of the MME based quickest spectrum sensing detector is subsequently given in \ref{sec:evaluation}. We conclude the paper in \ref{sec:conclusion} and discuss some future research directions.

\section{System Model}
\label{sec:system_model}
In this Section, we introduce the system model and fix notation. We introduce the model in general with $\numusers$ SUs. From \ref{sec:mme_distributions} on it will be confined to two SUs ($\numusers = 2$) only.

The basic hypothesis testing problem of detecting the presence of a PU with $\numusers$ collaborating SUs can be stated in the following form:
\begin{equation*}
\begin{aligned}
\hypzero: &~ \vec{\sample}(\timeidx) = \vec{\noise}(\timeidx) \\
\hypone: &~ \vec{\sample}(\timeidx) = \vec{\signal}(\timeidx) + \vec{\noise}(\timeidx) \;,    
\end{aligned}
\end{equation*}
where $\vec{\sample}(\timeidx)$ is a $\numusers \times 1$ vector of complex baseband samples collected by the SUs at time index $\timeidx \in \N = \{1, 2, 3, \dots\}$. Furthermore, the complex $\numusers \times 1$ vectors $\vec{\signal}(\timeidx)$ and $\vec{\noise}(\timeidx)$ stand for the PU's signal and additive noise, respectively. If $\hypzero$ is true the PU is not transmitting and the SUs receive noise only. If $\hypone$ is true, the PU is transmitting such that the SUs receive a noisy version of the PU signal. Combining $\numsamples$ samples into a $\numusers \times \numsamples$ sample matrix yields $\mat{\sample} = (\vec{\sample}(1), \vec{\sample}(2), \dots, \vec{\sample}(\numsamples))$. Similarly, a signal matrix $\mat{\signal}$ and a noise matrix $\mat{\noise}$ can be constructed, so it follows $\mat{\sample} = \mat{\noise}$ under $\hypzero$ and $\mat{\sample} = \mat{\signal} + \mat{\noise}$ under $\hypone$. Using this notation, we can calculate the sample covariance matrix as $\mat{\cov}_{\sample} = \frac{1}{\numsamples} \mat{\sample}\herm{\mat{\sample}}$.

In this work we consider a phase shift keying (PSK) signal subject to basic additive white Gaussian noise (AWGN). Thus, each column of the signal matrix $\mat{\signal}$ is assumed to have the form $\vec{\signal}_\idxvar = \vec{\signal}(\idxvar) = \sqrt{\snr} \, \sigsymbol_\idxvar \, \onevec_\numusers$ for $\idxvar = 1, \dots, \numsamples$. Here, $\snr$ is the signal-to-noise ratio (SNR), $\sigsymbol_\idxvar \in \C$ is a complex PSK symbol on the unit circle (i.e. $\abs{\sigsymbol_\idxvar} = 1$) and $\onevec_\numusers$ is a column vector of dimension $\numusers$ containing only ones. The PSK symbols $\sigsymbol_\idxvar$ sent by the PU are assumed to be i.i.d. and are drawn from a uniform distribution over an arbitrary PSK alphabet. The noise matrix is assumed to follow a jointly complex circularly symmetric Gaussian distribution, such that $\mat{\noise}\dist\SCN{\zeromat_{\numusers \times \numsamples}}{\I_\numusers\hadamard\I_\numsamples}$, where $\I_\numusers$ is the identity matrix of dimension $\numusers$. This means, that the real- and imaginary part of each matrix entry of $\mat{W}$ is independently Gaussian $\normal{0}{1/2}$ distributed and the matrix entries are mutually independent. Furthermore, we assume that the PU signal is independent of the noise. Under this model, every receiver is assumed to have the same SNR. This will be essential for the analytical derivation of the PDF under $\hypone$ in \ref{sec:mme_distributions}.

The test statistic of the well known MME detector is:
\begin{equation}
\label{eq:teststat}
\teststat = \frac{\eig_1}{\eig_\numusers},
\end{equation}
where $\eig_1 \geq \dots \geq \eig_\numusers$ are the ordered eigenvalues of the sample covariance matrix $\mat{\cov}_{\sample}$ \cite{zeng_maximum-minimum_2007, zeng_eigenvalue-based_2009}. Since scaling of the sample covariance matrix results in the same scaling of its eigenvalues, the ratio is not affected by it. Thus, we will omit the normalization factor and use the scaled sample covariance matrix $\mat{\cov} = \mat{\sample}\herm{\mat{\sample}}$ in the following. Similarly, the test statistic $\teststat$ is only dependent on the SNR $\snr$, but independent of the actual noise power under both hypotheses. Hence, our system model uses the SNR as a parameter directly, as knowledge of the actual noise power is unnecessary.

Under hypothesis $\hypzero$ the sample covariance matrix is simply $\mat{\cov}_0 = \mat{\noise}\herm{\mat{\noise}}$. This is a complex uncorrelated central Wishart matrix of dimension $\numusers$ with $\numsamples$ degrees of freedom \cite{james_distributions_1964}, which we denote as $\mat{\cov}_0 \dist \CWUC{\numusers}{\numsamples}$. 

Under hypothesis $\hypone$, the sample covariance matrix is

\begin{align}
\label{eq:h1_cov_mat}
\mat{\cov}_1 &= (\mat{\signal} + \mat{\noise})\herm{(\mat{\signal} + \mat{\noise})} \nonumber \\
&= \mat{\signal}\herm{\mat{\signal}} + \mat{\signal}\herm{\mat{\noise}} + \mat{\noise}\herm{\mat{\signal}} + \mat{\noise}\herm{\mat{\noise}} \;.
\end{align}
The first part of \ref{eq:h1_cov_mat} written using the columns $\matc{\signal}{\idxvar}$ of $\mat{\signal}$ gives

\begin{align}
\label{eq:h1_cov_mat_one}
\mat{\signal}\herm{\mat{\signal}} &= \sum_{\idxvar = 1}^{\numsamples} \matc{\signal}{\idxvar} \herm{(\matc{\signal}{\idxvar})} \nonumber \\
&= \snr \sum_{\idxvar = 1}^{\numsamples} \underbrace{\sigsymbol_\idxvar^{} \conj{\sigsymbol_\idxvar}}_{\abs{\sigsymbol_\idxvar}^2} \, \onevec_\numusers^{} \transp{\onevec_\numusers} = \snr \numsamples \, \onevec_\numusers^{} \transp{\onevec_\numusers} \;.
\end{align}
Evidently, \ref{eq:h1_cov_mat_one} is constant. Similarly, the middle parts of \ref{eq:h1_cov_mat} can be expanded as
\begin{equation*}
\mat{\signal}\herm{\mat{\noise}} = \sum_{\idxvar = 1}^{\numsamples} \matc{\signal}{\idxvar} \herm{(\matc{\noise}{\idxvar})} = \sum_{\idxvar = 1}^{\numsamples} \sqrt{\snr} \sigsymbol_\idxvar \, \onevec_\numusers^{} \herm{(\matc{\noise}{\idxvar})}
\end{equation*}
and 
\begin{equation*}
\mat{\noise}\herm{\mat{\signal}} = \sum_{\idxvar = 1}^{\numsamples} \matc{\noise}{\idxvar} \herm{(\matc{\signal}{\idxvar})} = \sum_{\idxvar = 1}^{\numsamples} \sqrt{\snr}  \conj{\sigsymbol_\idxvar} \, \matc{\noise}{\idxvar} \transp{\onevec_\numusers} \,,
\end{equation*}
respectively. Since the entries of the vectors $\matc{\noise}{\idxvar}$ are complex circularly symmetric Gaussian, their distribution is unaffected by complex rotations, which occur by multiplication with the unit circle PSK symbols $\sigsymbol_\idxvar$. Therefore the distribution of the middle part of \ref{eq:h1_cov_mat}, $\mat{\signal}\herm{\mat{\noise}} + \mat{\noise}\herm{\mat{\signal}}$, is identically to the distribution of $\sqrt{\snr} \onevec_\numusers^{} \transp{\onevec_\numusers} \herm{\mat{\noise}} + \sqrt{\snr} \mat{\noise} \onevec_\numusers^{} \transp{\onevec_\numusers}$. Hence, it makes no difference for the distribution of the sample covariance matrix (and thereby also no difference for the distribution of its eigenvalues) whether an arbitrary PSK signal (including 4QAM for that matter) or a constant signal is observed. Thus, $\mat{\cov}_1$ is a complex uncorrelated non-central Wishart matrix of dimension $\numusers$ with $\numsamples$ degrees of freedom, c.f. \cite{james_distributions_1964}. In short notation we write $\mat{\cov}_1 \dist \CWUN{\numusers}{\numsamples}{\mat{\ncm}}$. Here, the parameter $\mat{\ncm} = \expop[\mat{(\signal + \noise})]\herm{\expop[(\mat{\signal + \noise})]} =  \snr \numsamples \, \onevec_\numusers^{} \transp{\onevec_\numusers}$ is called the non-centrality matrix.

In order to determine the PDF of the test statistic $\teststat$ under hypothesis $\hypone$ in \ref{sec:mme_distributions}, we need the ordered eigenvalues $\eigncm_1 \geq \dots \geq \eigncm_K$ of the non-centrality matrix $\mat{\ncm}$. For the rank one matrix $\mat{\ncm}$ they are readily given as $\eigncm_1 = \snr \numusers \numsamples$, while $\eigncm_2 = \dots = \eigncm_K = 0$. 

\section{MME Distributions for $\hypzero$ \& $\hypone$}
\label{sec:mme_distributions}
As a special case, we have investigated the scenario of two cooperating users in more detail, so for the following $\numusers = 2$. For this case, the sample covariance matrices are distributed according to $\mat{\cov}_0 \dist \CWUC{2}{\numsamples}$ and $\mat{\cov}_1 \dist \CWUN{2}{\numsamples}{\snr \, \numsamples \, \onevec_2^{} \transp{\onevec_2}}$, respectively. It turns out, that for this simplification the PDFs under both hypotheses can be found explicitly.

\subsection{PDF under hypothesis $\hypzero$}
Under hypothesis $\hypzero$, the PDF of the test statistic $\teststat$ for $\numusers = 2$ has been found in \cite{kortun_performance_2011} as:

\begin{align}
\label{eq:scn_pdf_0_final}
\pdf_0(\pdfvar) & = \frac{\Gamma(2\numsamples)}{\Gamma(\numsamples)\Gamma(\numsamples-1)} \left(1-\frac{1}{\pdfvar}\right)^2 \left(\frac{1}{\pdfvar}\right)^\numsamples \left(1+\frac{1}{\pdfvar}\right)^{-2\numsamples} \nonumber \\
& = \frac{(\numsamples-1) \Gamma(2\numsamples)}{[\Gamma(\numsamples)]^2} \frac{(\pdfvar-1)^2 \, \pdfvar^{(\numsamples-2)}}{(\pdfvar+1)^{2\numsamples}},
\end{align}
for $ \pdfvar \ge 1$.

\subsection{PDF under hypothesis $\hypone$}
To develop the PDF of the test statistic $\teststat$ for $\numusers = 2$ under hypothesis $\hypone$, we follow the same way as \cite{matthaiou_analytical_2009}, where a version of the PDF was given for $\numsamples = 2$ and generalize this result to arbitrary $\numsamples$. We collect the ordered eigenvalues of $\mat{\cov}_1$ and $\mat{\ncm}$ in the vectors $\vec{\eig} = \transp{(\eig_1, \eig_2)}$ and $\vec{\eigncm} = \transp{(\eigncm_1, \eigncm_2)}$, respectively. Then, we can find the joint PDF of the ordered eigenvalues of $\mat{\cov}_1$ in \cite{matthaiou_condition_2010} as:
\begin{align*}
& \;\;\;\; \pdf_1^{\vec{\eig}}(\vec{\eig}) = \pdfconst_{un} \, \det{\vandmat{\vec{\eig}}} \, \det{\hypgeomat{\vec{\eig}}{\vec{\eigncm}}} \, \prod_{\idxvar = 1}^{2} \pdfconstexp{\eig_\idxvar} \\
& = \, \pdfconst_{un} \, e^{-(\eig_1 + \eig_2)} \, (\eig_1\!-\!\eig_2) \, (\eig_1\eig_2)^{(\numsamples-2)} \, \det{\hypgeomat{\vec{\eig}}{\vec{\eigncm}}}, \nonumber
\end{align*}
where $\det{\vandmat{\vec{\eig}}}$ is the determinant of a Vandermonde matrix built from $\vec{\eig}$, $\det{\hypgeomat{\vec{\eig}}{\vec{\eigncm}}}$ is the determinant of a matrix where the entry of the $\matridx$-th row and $\matcidx$-th column follows standard generalized hypergeometric functions $\hypgeozero{1}{\numsamples-1}{\eig_\matcidx \eigncm_\matridx}$ (see \cite{jeffrey_table_2007}, (9.14.1)), $\pdfconstexp{\eig_\idxvar} = \eig_\idxvar^{(\numsamples-2)} \, e^{-\eig_\idxvar}$ and
\begin{equation*}
\pdfconst_{un} = \frac{e^{-(\eigncm_1+\eigncm_2)}}{[\fac{(\numsamples-2)}]^2 \, (\eigncm_1 - \eigncm_2)} \;.
\end{equation*}

We are interested in the PDF of the test statistic $\teststat$, however, which is the ratio between the two ordered eigenvalues. By substituting $\eig_1 = \pdfvar \eig_2$ we can obtain the desired marginal PDF by applying the transformation in \ref{eq:scn_pdf_1_step1}.
\begin{bigequations}
\begin{equation}
\label{eq:scn_pdf_1_step1}
\pdf_1(\pdfvar) = \, \int_{0}^{\infty} \eig_2 \, \pdf_1^{\vec{\eig}}(\pdfvar \eig_2, \eig_2) \ud\eig_2 = \, \pdfconst_{un} \int_{0}^{\infty} e^{-\eig_2(\pdfvar+1)} \eig_2^2(\pdfvar-1) (\pdfvar\eig_2^2)^{(\numsamples-2)} \det{\hypgeomat{\transp{(\pdfvar \eig_2, \eig_2)}}{\vec{\eigncm}}} \ud\eig_2
\end{equation}

\begin{equation}
\label{eq:det_detail}
\det{\hypgeomat{\transp{(\pdfvar \eig_2, \eig_2)}}{\vec{\eigncm}}} = \, \hypgeozero{1}{\numsamples-1}{\pdfvar \eig_2 \eigncm_1} \, \hypgeozero{1}{\numsamples-1}{\eig_2 \eigncm_2} - 
\hypgeozero{1}{\numsamples-1}{\pdfvar \eig_2 \eigncm_2} \, \hypgeozero{1}{\numsamples-1}{\eig_2 \eigncm_1}
\end{equation}
\begin{align}
\label{eq:scn_pdf_1_step2}
& \pdf_1(\pdfvar) = \, \pdfconst_{un} \, [\fac{(\numsamples-2)}]^2 (\pdfvar-1) \pdfvar^{(\numsamples-2)} (\pdfvar \eigncm_1 \eigncm_2)^{-\left(\frac{\numsamples-2}{2}\right)} \, \times \\
& \int_{0}^{\infty} \eig_2^\numsamples e^{-\eig_2(\pdfvar+1)} \left[ \bessel{(\numsamples-2)}(2\sqrt{\pdfvar \eig_2 \eigncm_1}) \, \bessel{(\numsamples-2)}(2\sqrt{\pdfvar \eig_2 \eigncm_2}) -
\bessel{(\numsamples-2)}(2\sqrt{\pdfvar \eig_2 \eigncm_2}) \, \bessel{(\numsamples-2)}(2\sqrt{\pdfvar \eig_1 \eigncm_2}) \right] \ud\eig_2 \nonumber
\end{align}
\hrulefill
\end{bigequations}
Explicit calculation of the remaining determinant of \ref{eq:scn_pdf_1_step1} yields \ref{eq:det_detail}.
The hypergeometric function $\hypgeozero{1}{\dummypar+1}{\dummyparB}$ can also be written in terms of the $\dummypar$-th order modified Bessel function of the first kind $\bessel{\dummypar}(\dummyparB)$, c.f. \cite{jeffrey_table_2007}, as $\hypgeozero{1}{\dummypar+1}{\dummyparB} = \fac{\dummypar} \, \dummyparB^{-\frac{\dummypar}{2}} \, \bessel{\dummypar}(2 \sqrt{\dummyparB})$. Substituting \ref{eq:det_detail} into \ref{eq:scn_pdf_1_step1}, using the identity for the hypergeometric function and simplifying gives \ref{eq:scn_pdf_1_step2}.
Lacking an analytical solution of the integral in \ref{eq:scn_pdf_1_step2}, we use the series expansion of the $a$-th order modified Bessel function of the first kind $\bessel{\dummypar}(\cdot)$:
\begin{equation}
\label{eq:bessel_series}
\bessel{\dummypar}(\dummyparB) = \sum_{\idxvar = 0}^{\infty} \frac{1}{\fac{\idxvar} \, \Gamma(\idxvar + \dummypar + 1)} \left(\frac{\dummyparB}{2}\right)^{(2\idxvar + \dummypar)} \;.
\end{equation}
Substituting the Bessel functions by \ref{eq:bessel_series} in \ref{eq:scn_pdf_1_step2}, using that $(\sum_{\idxvar = 0}^{\infty} \dummypar_\idxvar) (\sum_{\idxvarB = 0}^{\infty} \dummyparB_\idxvarB) =  \sum_{\idxvar = 0}^{\infty} \sum_{\idxvarB = 0}^{\infty} \dummypar_\idxvar \dummyparB_\idxvarB$ and simplifying we arrive at \ref{eq:scn_pdf_1_step3}.
\begin{bigequations}
\begin{equation}
\label{eq:scn_pdf_1_step3}
\pdf_1(\pdfvar) = \, \pdfconst_{un} \, [\fac{(\numsamples-2)}]^2 (\pdfvar-1) \pdfvar^{(\numsamples-2)}
\sum_{\idxvar = 0}^{\infty} \sum_{\idxvarB = 0}^{\infty} \frac{\left( (\pdfvar \eigncm_1)^\idxvar \, \eigncm_2^\idxvarB - (\pdfvar \eigncm_2)^\idxvar \, \eigncm_1^\idxvarB \right)}{\fac{\idxvar} \, \fac{\idxvarB} \, \Gamma(\idxvar + \numsamples - 1) \, \Gamma(\idxvarB + \numsamples - 1)}
\int_{0}^{\infty} \!\!\! \eig_2^{(\idxvar + \idxvarB + 2\numsamples - 2)} e^{-\eig_2(\pdfvar+1)} \ud\eig_2
\end{equation}

\begin{equation}
\label{eq:scn_pdf_1_final}
\pdf_1(\pdfvar) = \, \frac{ e^{-(\eigncm_1+\eigncm_2)} \, (\pdfvar-1) \pdfvar^{(\numsamples-2)}}{(\eigncm_1-\eigncm_2)}
\sum_{\idxvar = 0}^{\infty} \sum_{\idxvarB = 0}^{\infty} \frac{\Gamma(\idxvar + \idxvarB + 2\numsamples - 1) \, \pdfvar^\idxvar \left( \eigncm_1^\idxvar \eigncm_2^\idxvarB - \eigncm_2^\idxvar \eigncm_1^\idxvarB \right)}{\fac{\idxvar} \, \fac{\idxvarB} \, \Gamma(\idxvar + \numsamples - 1) \, \Gamma(\idxvarB + \numsamples - 1) (\pdfvar+1)^{(\idxvar + \idxvarB + 2\numsamples - 1)}}
\end{equation}

\begin{equation}
\label{eq:scn_pdf_1_rank1}
\pdf_1^\snr(\pdfvar) = \, e^{-(2 \snr \numsamples)} \, (\pdfvar-1) \pdfvar^{(\numsamples-2)}
\sum_{\idxvar = 0}^{\infty} \frac{(\pdfvar^\idxvar - 1) \, \Gamma(\idxvar + 2\numsamples - 1) \, (2\snr\numsamples)^{(\idxvar-1)}}{\fac{\idxvar} \, \Gamma(\idxvar + \numsamples - 1) \, \Gamma(\numsamples - 1) \, (\pdfvar+1)^{(\idxvar + 2\numsamples - 1)}}
\end{equation}
\hrulefill
\end{bigequations}
The remaining integral in \ref{eq:scn_pdf_1_step3} can be found in \cite{jeffrey_table_2007} (3.351.3) as 
\begin{equation*}
\int_0^\infty x^\dummyparB \, e^{-\dummypar x} \ud x = \frac{\fac{\dummyparB}}{\dummypar^{(\dummyparB+1)}} = \frac{\Gamma(\dummyparB + 1)}{\dummypar^{(\dummyparB+1)}}
\end{equation*}
to obtain $\pdf_1(\pdfvar)$ for $\pdfvar \geq 1$ and $\eigncm_1 \neq \eigncm_2$ in \ref{eq:scn_pdf_1_final}. There, we use the identity 
$\Gamma(\dummypar) = \fac{(\dummypar-1)}$ for $\dummypar \in \N$. As already discussed above, the non-centrality matrix $\mat{\ncm}$ of the PSK signal has rank one and therefore $\eigncm_1 = 2 \snr \numsamples$ and $\eigncm_2 = 0$. Setting $\eigncm_2 = 0$ and $\eigncm_1 = 2 \snr \numsamples$ in \ref{eq:scn_pdf_1_final}, we gain a simpler version of the PDF of the test statistic $T$ under hypothesis $\hypone$ for $\pdfvar \geq 1$ in \ref{eq:scn_pdf_1_rank1}, which has the SNR $\snr$ as a parameter directly.

The PDF $\pdf_0(\pdfvar)$ must be a special case of the PDF $\pdf_1^\snr(\pdfvar)$ in the limit for vanishing SNR. Examining \ref{eq:scn_pdf_1_rank1} in the limit for $\snr \rightarrow 0$, the only term in the sum which is unequal to zero is when $\idxvar = 1$, and it follows:
\begin{equation}
\label{eq:scn_pdf_1_limit}
\lim_{\snr \rightarrow 0} \pdf_1^{\snr}(\pdfvar) = \frac{(\numsamples - 1) \Gamma(2\numsamples)}{[\Gamma(\numsamples)]^2} \, \frac{(\pdfvar-1)^2 \, \pdfvar^{(\numsamples-2)}}{(\pdfvar+1)^{2\numsamples}}  = \pdf_0(\pdfvar)\;.
\end{equation}

\begin{figure}[!hbt]
  \centering
  \setlength\figureheight{7cm}
  \setlength\figurewidth{9cm}
  \import{figures/}{pdf_plot.tex}
  \caption{Plot of $\pdf_0(\pdfvar)$ (noise only) and $\pdf_1^\snr(\pdfvar)$ for different values of the SNR~$\snr$ and $\numsamples = 500$.}
  \label{fig:pdf_plot}
\end{figure}
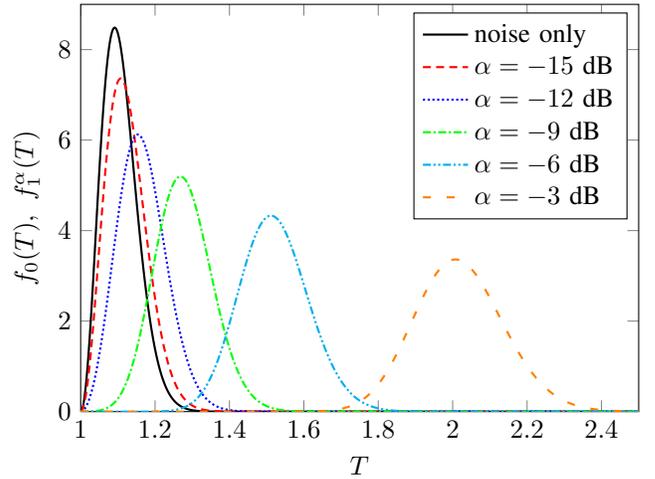

In \ref{fig:pdf_plot} the PDFs $\pdf_0(\pdfvar)$ and $\pdf_1^\snr(\pdfvar)$ are visualized, the latter for different values of the SNR $\snr$.

\subsection{Numerical Evaluation of the PDFs}
\label{sec:pdf_numa}
Numerical evaluation of the PDFs $\pdf_0(\pdfvar)$ and $\pdf_1^\snr(\pdfvar)$ is not straightforward, especially for large numbers of samples $\numsamples$, which are relevant in a spectrum sensing context. First, due to the factorials and exponents the numerical range needed quickly exceeds the range of the IEEE double precision format. Second, $\pdf_1^\snr(\pdfvar)$ involves an infinite sum and must therefore be approximated, e.g., by stopping the summation after a certain number of summands. One possibility is to utilize arbitrary-precision arithmetic to bypass numerical issues, however, the computation time is often unacceptably long. Thus, we will give a brief description of our implementation of the PDFs and compare them to empirical PDFs obtained by monte-carlo simulations.

The log-gamma function $\log(\Gamma(x))$ is often used for computations involving factorials or the gamma function itself. We used MATLAB for our numerical calculations, where an implementation of the log-gamma function is already provided. Therefore, we reformulate $\pdf_0(\pdfvar)$ to perform the numerically critical computations inside the exponential function:
\begin{align}
\label{scn_pdf_0_final_numa}
\pdf_0(\pdfvar) &= (\numsamples-1) \, (\pdfvar - 1)^2 \exp\left[ (\numsamples - 2) \, \log(\pdfvar)  \right. \nonumber \\ 
& \left. - 2 \numsamples \log(\pdfvar+1) + \log(\Gamma(2\numsamples))  \right. \\ 
& \left. - 2\log(\Gamma(\numsamples)) \right] \;. \nonumber 
\end{align}
In this way, we can evaluate the function for much higher $\numsamples$ than with a direct implementation of \ref{eq:scn_pdf_0_final}.

Since $\pdf_1^\snr(\pdfvar)$ contains an infinite sum, we approximate it by stopping the computation after a finite number of summands. Similarly to the approach of $\pdf_0(\pdfvar)$ we reformulate $\pdf_1^\snr(\pdfvar)$ to be able to use log-gamma functions for numerical computation:
\begin{equation}
\label{eq:scn_pdf_1_rank1_numa}
\pdf_1^\snr(\pdfvar) \approx (\pdfvar - 1) \sum_{\idxvar = 0}^{\sumapprox} \exp(\undummyf_\idxvar + \idxvar \log(\pdfvar)) - \exp(\undummyf_\idxvar),
\end{equation}
where
\begin{align*}
\undummyf_\idxvar & = \log(\Gamma(\idxvar+2\numsamples\!-1)) - \log(\Gamma(\idxvar+1)) - (2 \snr \numsamples) \\ 
& - \log(\Gamma(\numsamples-1)) + (\numsamples-2) \log(\pdfvar) \\
& - (\idxvar + 2\numsamples -1) \log(\pdfvar+1) + (\idxvar-1) \log(2 \snr \numsamples) \;.
\end{align*}

In \cite{matthaiou_analytical_2009}, where a version of $\pdf_1^\snr(\pdfvar)$ was developed with $\numsamples = 2$, it was reported that the number of summands can be chosen such that a further increase would only add a negligible amount (say $\leq 0.5~\%$) to the summation. However, for higher $\numsamples$ this criterion fails since the main contribution to the summation does not take place in the first summands anymore. To achieve a reasonable approximation, we check whether the integral over $\pdf_1^\snr(\pdfvar)$ is very close to one. In \ref{tab:pdf_1_rank1_numa}, we summarize values for $\sumapprox$ of \ref{eq:scn_pdf_1_rank1_numa}, such that the integral deviates less than $10^{-5}$ from one (using a trapezoidal quadrature rule with a bin width of $0.001$).
\begin{table}[htb]
\centering
\caption{Upper bound of summation ($\sumapprox$) to achieve a reasonable approximation of $\pdf_1^\snr$ using \ref{eq:scn_pdf_1_rank1_numa} for different values of the number of samples $\numsamples$ and the SNR $\snr$.}
\begin{tabular}{r||r|r|r|r|r}
$\numsamples \downarrow \, \backslash \, \overset{\rightarrow}{\snr}$ & -20 \dB & -15 \dB & -10 \dB & -5 \dB & 0 \dB \\ \hline \hline
50 & 10 & 20 & 30 & 60 & 200 \\
100 & 20 & 30 & 50 & 200 & 300 \\
500 & 30 & 60 & 200 & 400 & 2000 \\
1000 & 50 & 200 & 300 & 800 & 3000 \\
5000 & 200 & 400 & 2000 & 4000 & 20000 \\
10000 & 300 & 800 & 3000 & 7000 & 30000 \\
50000 & 2000 & 4000 & 20000 & 40000 & 200000 \\
\end{tabular}
\label{tab:pdf_1_rank1_numa}
\end{table}

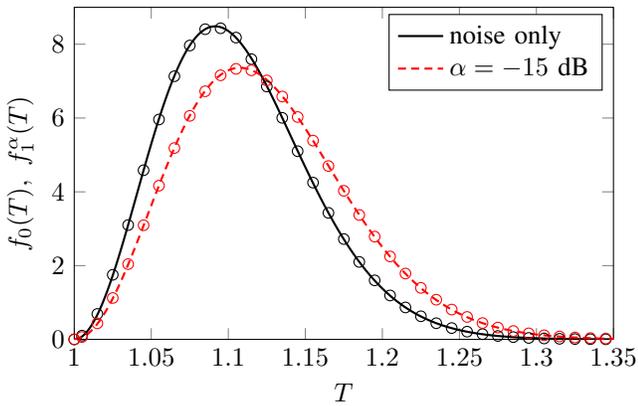
\begin{figure}[!ht]
  \centering
  \setlength\figureheight{6cm}
  \setlength\figurewidth{8.75cm}
  \import{figures/}{pdf_verification_plot.tex}
  \caption{Plot of $\pdf_0(\pdfvar)$ (noise only) and $\pdf_1^\snr(\pdfvar)$ for $\snr = -15~\dB$ and $\numsamples = 500$. Circles indicate values taken from an empirical PDF obtained by simulation.}
  \label{fig:pdf_verification_plot}
\end{figure}

To verify our implementation including our approximations, the PDFs $\pdf_0(\pdfvar)$ and $\pdf_1(\pdfvar)$ are plotted in \ref{fig:pdf_verification_plot} by evaluating \ref{scn_pdf_0_final_numa} and \ref{eq:scn_pdf_1_rank1_numa}, respectively. There, the values of the empirical PDFs obtained by digital simulation are drawn in circles. 

\section{Application: Block Detection}
\label{sec:BD}
In this Section, we apply the results from \ref{sec:mme_distributions} to a block detection scenario, before we turn to the quickest detection problem in the next Section. For a block detection scheme using the MME detector, the number of samples $\numsamples$ used in calculating the sample covariance matrix and the decision threshold $\threshold$ need to be determined. Intuitively, higher $\numsamples$ result in less overlap between the PDFs $\pdf_0(\pdfvar)$ and $\pdf_1^\snr(\pdfvar)$. Thus, choosing $\numsamples$ presents a trade-off between the detection performance and the time to reach a decision. The threshold $\threshold$, poses a trade-off between the probability of detection $\pd$ and the probability of false alarm $\pfa$: 
\begin{align}
& \pfa(\threshold) = 1 - \cdf_0(\threshold) = 1 - \int_1^{\threshold} \pdf_0(\pdfvar) \ud\pdfvar \; , \nonumber \\
& \pd(\threshold) = 1 - \cdf_1^\snr(\threshold) = 1 - \int_1^{\threshold} \pdf_1^\snr(\pdfvar) \ud\pdfvar \; . \label{eq:p_d}
\end{align}
As mentioned above, test statistic $\teststat$ of the MME detector is not directly dependent on the actual noise and signal powers, but rather on the SNR $\snr$. Typically, the SNR $\snr$ is assumed to be unknown beforehand. The decision threshold $\threshold$ is then chosen, such that the detector has false alarm rate $\pfa$.

Note, that it is assumed in block detection that no change of the hypothesis happens within a block. Consequently, when the CDFs under hypotheses $\hypzero$ and $\hypone$ are both known, the receiver operator characteristic (ROC) can be calculated theoretically. Thereby, the design of a detector is simplified significantly, as the detection performance can be determined beforehand without the need for simulations or empirical tests. 

The CDF under hypothesis $\hypzero$, i.e. $\cdf_0(\cdfvar)$ was given in \cite{matthaiou_condition_2010} as:
\begin{equation*}
\cdf_0(\cdfvar) = \pdfconst_{uc} \, (\ucdummycf(\cdfvar) - \ucdummycf(1)) \;,
\end{equation*}
where $\cdfvar \ge 1$,
\begin{align*}
\ucdummycf(\cdfvar) = &~\ucdummyf_1(\numsamples, \numsamples - 1, \cdfvar) - 2\ucdummyf_1(\numsamples - 1, \numsamples, \cdfvar) \\ 
&~+ \ucdummyf_1(\numsamples - 2, \numsamples + 1, \cdfvar) \; 
\end{align*}
and
\begin{equation*}
\ucdummyf_1(\ucdummyfpOne,\ucdummyfpTwo,\ucdummyfpThree) = \fac{(\ucdummyfpTwo - 1)} \left( \fac{\ucdummyfpOne} - \sum_{\idxvar = 0}^{(\ucdummyfpTwo - 1)} \frac{\fac{(\ucdummyfpOne + \idxvar)} \, \ucdummyfpThree^{\idxvar}}{\fac{\idxvar} \, (\ucdummyfpThree+1)^{\ucdummyfpOne+\idxvar+1}} \right) \;.
\end{equation*}

\begin{figure}[!ht]
  \centering
  \setlength\figureheight{6cm}
  \setlength\figurewidth{8.7cm}
  \import{figures/}{roc_plot.tex}
  \caption{ROC of the MME block detection scheme with $\numsamples = 500$ and various SNRs~$\snr$. Circles indicate values obtained from digital simulation. Compare also the corresponding PDFs in \ref{fig:pdf_plot}.}
  \label{fig:roc_plot}
\end{figure}
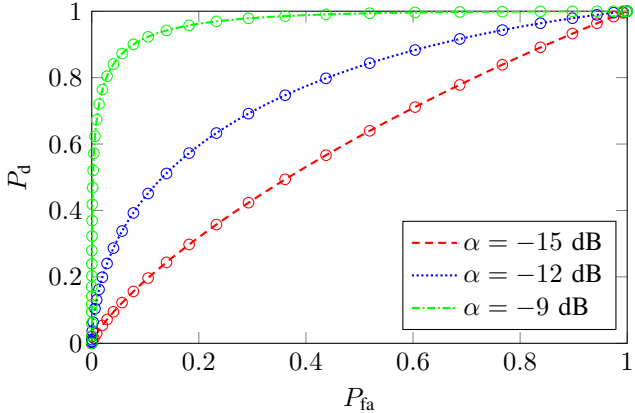

We are not aware of an analytical solution to the integral in \ref{eq:p_d}. However, with the help of the PDF $\pdf_1^\snr(\pdfvar)$ from \ref{eq:scn_pdf_1_final} we can numerically evaluate the integral to get $\pd$. Plotting $\pd$ over $\pfa$ for various thresholds results in the ROC. As an example, the ROC for $\numsamples = 500$ with various SNRs $\snr$ is shown in \ref{fig:roc_plot}. There, results from a digital simulation are indicated by circles to verify the theoretical ROC.

\section{Application: Quickest Detection}
\label{sec:QD}
This Section applies the results from \ref{sec:mme_distributions} to introduce quickest detection (QD) based on the eigenvalues of the sample covariance matrix. In contrast to block detection, where the objective is to determine which hypothesis is true, in QD it is assumed that it is known which hypothesis is true and that a change to the other hypothesis shall be detected with as little delay as possible. Contrary to evaluating the probabilities of detection $\pd$ and false alarm $\pfa$ as done in block detection, the design criterions for QD algorithms are the mean time to detection $\td$ and the mean time to false alarm $\tfa$. A well known algorithm used in QD is the cumulative sum (CUSUM) algorithm \cite{page_continuous_1954}. It is the optimal QD algorithm, in the sense that it achieves minimal $\td$ while satisfying $\tfa \ge c$, for any $c$ \cite{moustakides_optimal_1986}. Without loss of generality, we assume in the following that before the change time $\tchange$ hypothesis $\hypzero$ is true and at time $\tchange$ transmission of the PU starts, i.e., hypothesis $\hypone$ becomes true (see also \ref{fig:QD}). This can be formalized as follows. 

Let $\cusumseq_\timeidx$ be a sequence of random variables, which for each $\timeidx$ is independently distributed according to the PDF $\pdf_{\cusumpdfpar_0}(\cusumpdfvar)$ with parameter $\cusumpdfpar_0$ before the change time, i.e. for $\timeidx < \tchange$ and according to the PDF $\pdf_{\cusumpdfpar_1}(\cusumpdfvar)$ with parameter $\cusumpdfpar_1$ at and after the change time, i.e. for $\timeidx \ge \tchange$. Samples  $\cusumsample(\timeidx)$ are taken in order to detect the change from $\hypzero \rightarrow \hypone$.
The CUSUM algorithm relies on the log-likelihood ratio of the received sample at time index $\timeidx$: $\llr_\cusumsample(\timeidx) = \log\left(\pdf_{\cusumpdfpar_1}(\cusumsample(\timeidx)) / \pdf_{\cusumpdfpar_0}(\cusumsample(\timeidx))\right)$. Over time, $\llr_\cusumsample(\timeidx)$ shows a positive drift if $\hypone$ is true and a negative drift if $\hypzero$ is true. A cumulative sum of the log-likelihood ratios of consecutive samples is formed by the CUSUM algorithm. However, only those consecutive samples are used for the summation, which result in the largest sum. In other words, only the time interval showing a consistent positive drift is considered for summation. A convenient recursive formulation of the CUSUM algorithm can be given as \cite{page_continuous_1954, basseville_detection_1993}:
\begin{align}
    \label{eq:cusum_theo}
    \cusum_\cusumsample(\timeidx) &= \max_{0 \leq \glrvar \leq \timeidx} \sum_{\idxvar = \glrvar + 1}^{\timeidx} \log\left(\frac{\pdf_{\cusumpdfpar_1}(\cusumsample(\idxvar))}{\pdf_{\cusumpdfpar_0}(\cusumsample(\idxvar))}\right) \nonumber \\
    &= \ppart{\cusum_\cusumsample(\timeidx-1) + \llr_\cusumsample(\timeidx)},
\end{align}
where the initial value is defined as $\cusum_\cusumsample(0) = 0$ and the positive part is defined as $\ppart{\cdot} = \max(\cdot, 0)$. At each time index $\timeidx$ the value $\cusum_\cusumsample(\timeidx)$ is compared to a threshold $\threshold > 0$ and if $\cusum_\cusumsample(\timeidx) > \threshold$ the algorithm decides that a change to $\hypone$ has happened. As can be seen from \ref{eq:cusum_theo}, $\cusum_\cusumsample(\timeidx)$ may never reach negative values. Since $\llr_\cusumsample(\timeidx)$ shows a negative drift under $\hypzero$ and we are interested in detecting a potential change from $\hypzero \rightarrow \hypone$, accumulating negative values would result in a much longer time to reach the threshold after the change time $\tchange$ and thus in a larger mean time to detection $\td$. Note, that the parameters $\cusumpdfpar_0$ and $\cusumpdfpar_1$ must be known to use the CUSUM algorithm. If the parameters are unknown, the algorithm can be adapted to use the generalized log-likelihood ratio. We will study the application of QD using the MME test statistic with and without knowledge of the SNR in the following.

\subsection{Known SNR}
First, we study the model from \ref{sec:system_model} for the case that the SNR $\snr$ is known to the receiver. To adapt the results from \ref{sec:mme_distributions}, we define a time-dependent version of the test statistic $\teststat$ from \ref{eq:teststat}. Introducing a block index $\blockidx \in \N$, we define $\vec{\eig}(\blockidx)$ as the vector of ordered eigenvalues of the scaled sample covariance matrix of block $\blockidx$: $\mat{\cov}(\blockidx) = \mat{\sample}(\blockidx)\herm{\mat{\sample}}(\blockidx)$, where $\mat{\sample}(\blockidx) = \left(\vec{\sample}((\blockidx - 1) \numsamples + 1),~\cdots, \vec{\sample}(\blockidx \numsamples) \right)$. In other words, $\vec{\eig}(\blockidx)$ contains the eigenvalues of the $\blockidx$-th consecutive sample covariance matrix built from a block of $\numsamples$ non-overlapping samples. Thus, since we assume $\numusers = 2$, the test statistic follows as:
\begin{equation*}
    \teststat(\blockidx) = \frac{\eig_1(\blockidx)}{\eig_2(\blockidx)}\;.
\end{equation*}

To take over existing results from QD, we make an additional assumption, which is similarly present in block detection: $\tchange \in \{(\blockidx - 1) \numsamples + 1~|~\blockidx \in \N\}$. This means that no hypothesis change may happen within a block. Depending on whether $\hypzero$ or $\hypone$ is true, $\teststat(\blockidx)$ is distributed according to $\pdf_0(\pdfvar)$ or $\pdf_1^{\snr}(\pdfvar)$, respectively. As shown in \ref{eq:scn_pdf_1_limit}, $\lim\limits_{\snr \rightarrow 0} \pdf_1^{\snr}(\pdfvar) = \pdf_0(\pdfvar)$ so the only parameter influenced by the change is the SNR $\alpha$.

The log-likelihood ratio can be found by inserting \ref{eq:scn_pdf_0_final} and \ref{eq:scn_pdf_1_final} into its definition and simplifying as seen in \ref{eq:llr}.
\begin{bigequations}
\begin{equation}
\label{eq:llr}
    \llr(\blockidx) = \log\left(\frac{\pdf_1^\snr(\teststat(\blockidx))}{\pdf_0(\teststat(\blockidx))}\right)
                    = \log\left(\frac{e^{-2\snr\numsamples}}{(\teststat(\blockidx) - 1)} \sum_{\idxvar = 0}^{\infty} \frac{(\teststat(\blockidx)^{\idxvar} - 1)\, (2\snr\numsamples)^{(\idxvar-1)} \, \pochhammer{2\numsamples}{(\idxvar-1)}}{\fac{\idxvar} \, (\teststat(\blockidx) + 1)^{(\idxvar-1)} \, \pochhammer{\numsamples}{(\idxvar-1)}} \right)
\end{equation}
\hrulefill
\end{bigequations}
There, we use the Pochhammer symbol defined as $\pochhammer{a}{b} = \fac{(a+b-1)}/\fac{(a-1)}$. Note, that \ref{eq:llr} can be evaluated numerically using a similar approach to the one used for the PDFs in \ref{sec:pdf_numa}. Using the log-likelihood ratio, we can give the CUSUM algorithm for our model as:
\begin{equation}
\label{eq:cusum}
    \cusum(\blockidx) = \ppart{\cusum(\blockidx-1) + \llr(\blockidx)} \;.
\end{equation}

When designing the CUSUM detector, a method for finding a suitable threshold $\threshold$ is desirable. Since the exact calculation of the mean time to detection $\td$ and the mean time to false alarm $\tfa$ is very complicated in general, we turn to bounds already studied in the literature. Of special interest are an upper bound on $\td$ and a lower bound on $\tfa$ to ease the design of the threshold $\threshold$. Note, that we are evaluating the bounds using the timescale of the block index $\blockidx$. To evaluate the bounds on a sample based timescale (time index $\timeidx$ in \ref{sec:system_model}), the bounds must be multiplied by the number of samples $\numsamples$ used for calculating the sample covariance matrix in each block.

Starting with the upper bound on $\td$, we find from \cite{wald_sequential_1947, basseville_detection_1993}, that
\begin{equation}
    \label{eq:cusum_td_ub}
    \td \leq \frac{(\threshold + \cusumsup(\pdf_1^\snr))}{\expop_{\pdf_1^\snr}\left[\llr(\blockidx)\right]} \;, 
\end{equation}
where 
\begin{equation*}
    \cusumsup(\pdf) = \sup_{\cusumsupshift > 0} \; \expop_\pdf\left[\llr(\blockidx) - \cusumsupshift ~|~ \llr(\blockidx) \geq \delta > 0\right] \;.
\end{equation*}
Here, we denote the expectation over the PDF $\pdf(\pdfvar)$ as $\expop_{\pdf}[\cdot]$. Likewise, a lower bound on $\tfa$ can be found in \cite{basseville_detection_1993}:
\begin{equation}
    \label{eq:cusum_tfa_lb_orig}
    \tfa \geq \frac{1}{\expop_{\pdf_0}\left[\llr(\blockidx)\right]} \, \left(\frac{e^{-\cusumroot\threshold} - 1}{\cusumroot} + \threshold + \cusumsup(\pdf_0) \right) \;, 
\end{equation}
where $\cusumroot < 0$ is the single non-zero root of $\expop_{\pdf_0}[e^{-\cusumroot\,\llr(\blockidx)}] = 1$, which can be found by solving:
\begin{equation}
    \label{eq:cusumroot}
    \expop_{\pdf_0}[e^{-\cusumroot\,\llr(\blockidx)}] = \int_1^{\infty} \left(\frac{\pdf_1^\snr(\pdfvar)}{\pdf_0(\pdfvar)}\right)^{-\cusumroot} \pdf_0(\pdfvar) \ud\pdfvar \overset{!}{=} 1 \;.
\end{equation}
Obviously, $\cusumroot = -1$ solves \ref{eq:cusumroot}, since only the integral over $\pdf_1^\snr(\pdfvar)$ remains. Thus, \ref{eq:cusum_tfa_lb_orig} can easily be simplified to:
\begin{equation}
    \label{eq:cusum_tfa_lb}
    \tfa \geq \frac{1}{\expop_{\pdf_0}\left[\llr(\blockidx)\right]} \, \left(1 - e^{\threshold} + \threshold + \cusumsup(\pdf_0) \right) \;.
\end{equation}

Neither $\expop_{\pdf_0}\left[\llr(\blockidx)\right]$, $\expop_{\pdf_1^\snr}\left[\llr(\blockidx)\right]$, $\cusumsup(\pdf_0)$ nor $\cusumsup(\pdf_1^\snr)$ lend themselves to proper analytical treatment, but can be calculated numerically.

A second lower bound on $\tfa$ reported in the literature (see e.g., \cite{basseville_detection_1993}) is simply 
\begin{equation}
\label{eq:cusum_tfa_bound_simple}
\tfa \ge e^{\threshold} \;.
\end{equation}
Although this bound is very simple to compute, we found that for our model and the threshold values we considered in the numerical evaluation, this bound turns out to be very loose. Since \ref{eq:cusum_tfa_lb} is much tighter to the value of $\tfa$ obtained by simulation, we will omit \ref{eq:cusum_tfa_bound_simple} in our numerical evaluation.

\subsection{Unknown SNR}
A more realistic scenario is that the SNR is unknown to the receiver beforehand. In this case, the CUSUM algorithm cannot be used directly. However, a similar algorithm based on the 
generalized likelihood ratio test (GLRT) can be utilized instead. The GLRT allows to perform likelihood ratio tests when the parameters of the involved PDFs are unknown, by estimating these parameters using a maximum likelihood estimation (MLE) beforehand \cite{kay_fundamentals_1998}. In our scenario, the only unknown parameter is the SNR and the GLRT for this case can be given as:
\begin{equation}
\label{eq:glrt}
\glrt(\blockidx) = \log\left( \sup_{\glrsnr} \frac{\pdf_1^\glrsnr(\teststat(\blockidx))}{\pdf_0(\teststat(\blockidx))} \right) \;,        
\end{equation}
where $\glrsnr$ is the estimated SNR. Utilizing the GLRT test on an i.i.d. sequence of samples, the generalized likelihood ratio (GLR) algorithm can be constructed as \cite{basseville_detection_1993}:
\begin{align}
\label{eq:glr}
\glr(\blockidx) &= \max_{0 \leq \glrvar \leq \blockidx} \sup_{\glrsnr} \log\left( \prod_{\idxvar = \glrvar + 1}^{\blockidx} \frac{\pdf_1^\glrsnr(\teststat(\idxvar))}{\pdf_0(\teststat(\idxvar))} \right) \nonumber \\
&= \max_{0 \leq \glrvar \leq \blockidx} \sup_{\glrsnr} \sum_{\idxvar = \glrvar + 1}^{\blockidx} \log\left( \frac{\pdf_1^\glrsnr(\teststat(\idxvar))}{\pdf_0(\teststat(\idxvar))} \right) \;.
\end{align}

We are not aware of an analytical form of the supremum in \ref{eq:glrt}, therefore we evaluate it numerically. Note also, that the GLR algorithm of \ref{eq:glr} cannot be written in a recursive form. In contrast to the CUSUM algorithm, the GLR algorithm requires memory of all previous samples. 
This is due to the numerical evaluation of the supremum and the subsequent maximization in \ref{eq:glr}, which also has a higher computational complexity. 

\begin{figure}[!ht]
  \centering
  \setlength\figureheight{6cm}
  \setlength\figurewidth{8.7cm}
  \import{figures/}{QD_t_d_plot.tex}
  \caption{Theoretical upper bound from \ref{eq:cusum_td_ub} on the mean time to detection $\td$ of the CUSUM algorithm. Simulated performance of both the CUSUM and GLR algorithm is shown for $\numsamples = 500$ and SNR $\snr = -15~\dB$. The GLR numerically evaluates the supremum over $\glrsnr$ within the interval $[-20~\dB, -5~\dB]$ in $0.1~\dB$ steps.}
  \label{fig:QD_t_d_plot}
\end{figure}
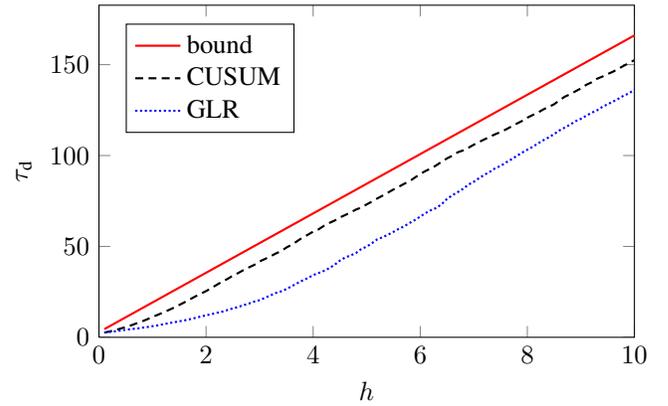

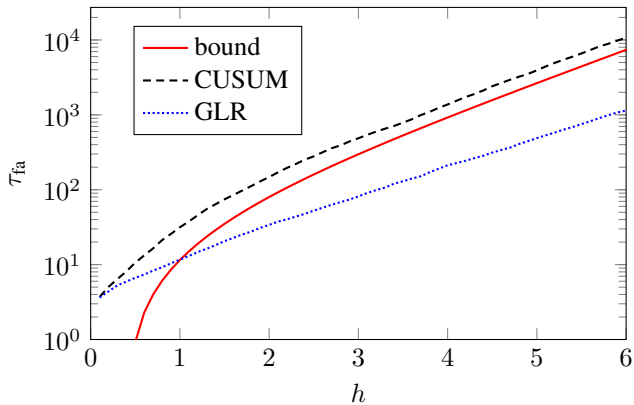
\begin{figure}[!ht]
  \centering
  \setlength\figureheight{6cm}
  \setlength\figurewidth{8.7cm}
  \import{figures/}{QD_t_fa_plot.tex}
  \caption{Theoretical lower bound from \ref{eq:cusum_tfa_lb} on the mean time to false alarm $\tfa$ of the CUSUM algorithm. Simulated performance of both the CUSUM and GLR algorithm is shown for $\numsamples = 500$ and SNR $\snr = -15~\dB$. The GLR numerically evaluates the supremum over $\glrsnr$ within the interval $[-20~\dB, -5~\dB]$ in $0.1~\dB$ steps. Note the logarithmic scaling of the ordinate.}
  \label{fig:QD_t_fa_plot}
\end{figure}

In \ref{fig:QD_t_d_plot} the mean time to detection $\td$ obtained from digital simulation is plotted over the threshold for the CUSUM and the GLR algorithm. Also, the theoretical upper bound on $\td$ from \ref{eq:cusum_td_ub} is depicted. Note, that the upper bound is only valid for the CUSUM algorithm. Similarly, in \ref{fig:QD_t_fa_plot} the mean time to false alarm $\tfa$ obtained by digital simulation is shown for both the CUSUM and the GLR algorithm with a logarithmic scaling of the ordinate. Again, the theoretical lower bound on $\tfa$ from \ref{eq:cusum_tfa_lb}, which is only valid for the CUSUM algorithm, is evaluated there. The values of $\td$ and $\tfa$ shown in the Figures \origref{fig:QD_t_d_plot} and \origref{fig:QD_t_fa_plot} are evaluated on a block-wise timescale, as already explained above.

\begin{figure}[!ht]
  \centering
  \setlength\figureheight{5cm}
  \setlength\figurewidth{7.8cm}
  \import{figures/}{glr_initial_effects.tex}
  \caption{Comparison of a typical run of the CUSUM and the GLR algorithm with $\numsamples = 500$ and SNR $\snr = -15~\dB$. The internal parameters $\glrsnr$ and $\glrvar$ of the GLR algorithm are shown in the bottom plot.}
  \label{fig:glr_initial_effects}
\end{figure}
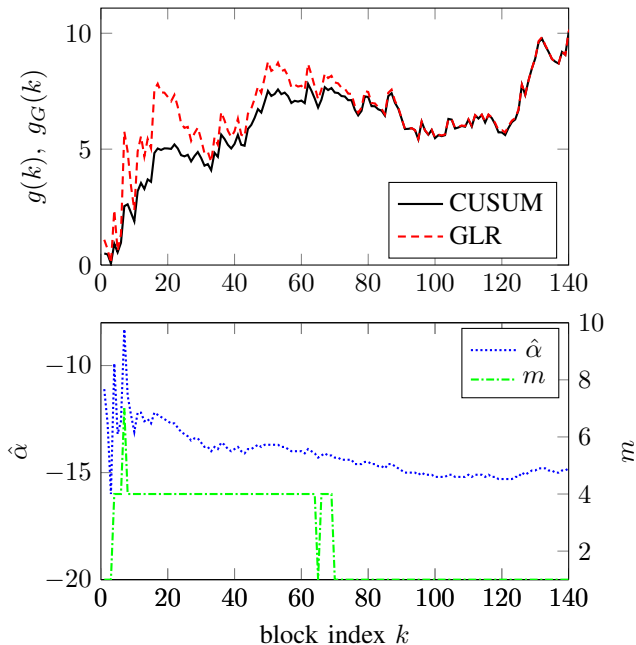

As can be seen in both \ref{fig:QD_t_d_plot,fig:QD_t_fa_plot}, the GLR algorithm shows a more aggressive behavior compared to the CUSUM algorithm. This can be explained by two effects, predominantly present at the beginning of the algorithm run-time. Firstly, the SNR estimate $\glrsnr$ used by the GLR algorithm \ref{eq:glr} may over- or underestimate the SNR. If, for instance, the first couple of sample covariance matrices result in a greater than average test statistic $\teststat(\blockidx)$ for the respective SNR, the GLR will overestimate the SNR $\glrsnr$ and hence the log-likelihood ratio will be greater than it would be for the true SNR $\snr$. Secondly, when the GLR algorithm \ref{eq:glr} determines its parameter $\glrvar$, it may also choose larger values for $\glrvar$ than it would if the SNR estimation $\glrsnr$ had returned the true SNR $\snr$. In \ref{fig:glr_initial_effects}, a typical run of both the CUSUM and the GLR algorithm are depicted. The bottom of \ref{fig:glr_initial_effects} shows the internal parameters of the GLR algorithm. Both effects can be observed as discussed above. With advancing run-time, the SNR estimation $\glrsnr$ converges to the true SNR $\snr$. Consequently, also the behavior of the GLR algorithm more closely resembles the behavior of the CUSUM algorithm. If the SNR estimation of the GLR algorithm returns the true SNR, both algorithms behave identically, as can be seen by comparing \ref{eq:cusum} and \ref{eq:glr}.

\section{Numerical Evaluation}
\label{sec:evaluation}
\begin{figure}[!ht]
  \centering
  \setlength\figureheight{7cm}
  \setlength\figurewidth{8.3cm}
  \import{figures/}{glr_performance.tex}
  \caption{Performance of the GLR algorithm, evaluated as $\pd$ and $\pfa$ over the algorithm run-time for the threshold $\threshold = 4.5$. The gray circular markers and the thin solid lines indicate the performance of the MME block detector designed for a block length of $10^5$ samples and $\pfa = 0.0147$, which results in $\pd = 0.9269$ at SNR $\snr = -20~\dB$.}
  \label{fig:glr_performance}
\end{figure}
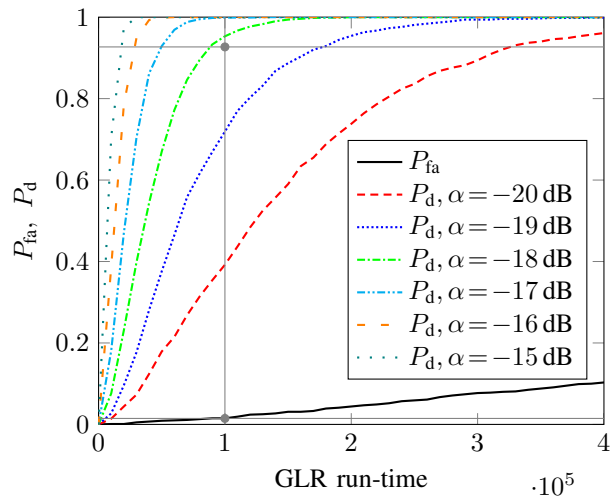

A fair comparison between the GLR algorithm, which is a quickest detection algorithm and the MME block detection scheme is difficult to obtain due to the assumptions present in the two different detection problems. In the following evaluation we design the MME block detector with a block length of $10^5$ samples and probability of false alarm $\pfa = 0.0147$. Using the results from \ref{sec:BD}, we find the probability of detection $\pd = 0.9269$ for the SNR $\alpha = -20~\dB$.
When using the GLR algorithm, a run-time has to be decided because the algorithm will eventually generate a false alarm. Evidently, this run-time is to be chosen such that the mean time to false alarm $\tfa$ is significantly higher than the algorithm run-time. We performed a monte-carlo simulation with the GLR algorithm for 1000 random seeds, with $\numsamples = 10000$ samples in each block and a variety of threshold values. Furthermore, two versions of the algorithm per monte-carlo instance are computed. The first one receives samples under hypothesis $\hypzero$, i.e. it only receives noise. The second one receives samples under hypothesis $\hypone$, so it receives signal with additive noise. For the detection performance this is a worst-case analysis, since the GLR algorithm is initialized with the value zero. If the hypothesis would switch from $\hypzero$ to $\hypone$ during the simulation, it might detect a longer consecutive positive drift due to the effects of the noise, resulting in potentially quicker detection. To calculate probabilities of false alarm and detection for the GLR algorithm, respectively, we consider different run-times and calculate how many of the 1000 monte-carlo instances have raised a false-alarm / have correctly detected the signal for a given run-time, threshold $\threshold$ and SNR $\snr$. Thus, we can compare the detection performance of the GLR algorithm to the fixed performance of the MME block detector. For the numerical evaluation of the supremum in \ref{eq:glr}, an SNR range of $[-25, -5]~\dB$ was evaluated in $0.1~\dB$ steps. In \ref{fig:glr_performance}, the GLR performance is depicted for several SNRs. The threshold $\threshold = 4.5$ was chosen, such that at a run-time of $10^5$ samples, i.e., the block size of the MME detector, the probabilities of false alarm are approximately equal (GLR: $\pfa = 0.0140$ vs. MME: $\pfa = 0.0147$).
As expected, the GLR quickest detection algorithm is able to detect higher SNRs much earlier than the MME detector at a lower or comparable false-alarm rate. However, at low SNRs the MME detector is faster. When considering realistic detection scenarios, the PUs SNR cannot be known beforehand. So the MME detector is designed to have a specific performance for the lowest SNR that has to be reliably detected. Evidently, in practice it would be more precise to consider an SNR interval, say e.g., $[-5, -20]~\dB$. If no further knowledge is available, we may model the SNR with a uniform distribution over said interval. Here, we see the advantage of the quickest detection approach. For a wide range of SNRs, it offers faster detection at comparable false-alarm performance, thus wasting less time on the detection process. On the one hand, this leads to  higher efficiency in using transmission opportunities, thereby increasing transmission rates. On the other hand, interference for the PU may be decreased by diminishing the reaction time to free the channel when the PU initiates a communication. As can be seen in \ref{fig:glr_performance}, there is a gap of approximately $2~\dB$ in the low SNR performance of the GLR quickest detection algorithm compared to the MME block detector. This is due to the fact that the GLR algorithm does not use the entirety of samples, but rather blocks of $\numsamples = 10000$ samples to calculate the sample covariance matrices. Thus, in choosing the number of samples $\numsamples$, one has to trade-off the detection delay for higher SNRs and the performance gap compared to the MME block detector for low SNRs. However, it must be stressed that this comparison is not completely fair, since the GLR algorithm introduced in \ref{eq:glr} essentially throws data away. This stems from the CUSUM analysis, where the bounds rely on the assumption that the samples of the test statistic used in the CUSUM algorithm are i.i.d.. Further improvements in this direction are discussed in \ref{sec:conclusion}.

\begin{figure}[!ht]
  \centering
  \setlength\figureheight{7cm}
  \setlength\figurewidth{8.3cm}
  \import{figures/}{glr_threshold_spread.tex}
  \caption{Performance of the GLR algorithm, evaluated as $\pd$ and $\pfa$ over the algorithm run-time for different thresholds $\threshold$ at SNR $\snr = -17~\dB$.}
  \label{fig:glr_threshold_spread}
\end{figure}
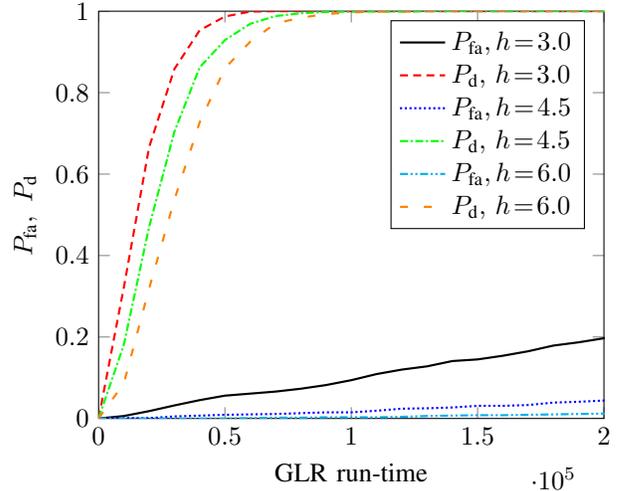

As for all detection algorithms the choice of the detection threshold has a huge impact on the detection performance. This effect can already be seen in \ref{fig:QD_t_d_plot,fig:QD_t_fa_plot}. To visualize the performance trade-off between $\pd$ and $\pfa$ directly, three different choices of the threshold $\threshold$ are compared in \ref{fig:glr_threshold_spread}.

\section{Conclusion \& Future Work}
\label{sec:conclusion}
In this work, we have investigated the potential for using quickest detection based on the sample covariance matrix for spectrum sensing applications. First, we have specified a simple phase shift keying (PSK) model with additive white Gaussian noise (AWGN), consisting of one primary user (PU) and $\numusers$ secondary users (SUs). Second, we characterized the Wishart distributions of the eigenvalues of the sample covariance matrix under both detection hypotheses $\hypzero$ (noise only) and $\hypone$ (PU signal + noise). Third, we derived analytical formulations of the PDF of the maximum-minimum eigenvalue (MME) test statistic, which is also called the standard condition number (SCN) of a matrix, under both hypotheses $\hypzero$ and $\hypone$ for the special case $\numusers = 2$. These PDFs have the number of samples $\numsamples$ in the sample covariance matrix and the signal-to-noise ratio $\snr$ as parameters. Then, we discussed the numerical computation of said PDFs. We put these results to application by explicitly calculating the ROC for the well known MME block detector without the need of expensive simulations. Subsequently, the main application of these results is discussed: quickest detection based on the sample covariance matrix, i.e. eigenvalue-based quickest detection for spectrum sensing. Two quickest detection algorithms were introduced, which rely on the MME test statistic. The CUSUM algorithm is applicable when the SNR of the PU signal is known and a GLR algorithm was deduced to cope with the situation when the SNR is unknown. Bounds on the mean time to false-alarm $\tfa$ and the mean time to detection $\td$ have been provided for the CUSUM algorithm. Numerical simulations illustrate the potential advantages of the quickest detection approach compared to the block detection scheme for spectrum sensing applications. It turns out, that for a wide range of SNRs the introduced GLR quickest detection algorithm offers faster detection at better or comparable false-alarm performance than the MME block detector.

Future research may focus on finding similar bounds on $\td$ and $\tfa$ for the GLR algorithm. Also, at very low SNRs the block detector is still faster than the GLR algorithm. This results from the fact that the GLR algorithm calculates its sample covariance matrices block-wise and hence uses less data than the block detector. One possibility of improvement would be to get rid of the block structure introduced in \ref{sec:QD} and perform updates of the samples covariance matrix on a sample basis. However, this would introduce stochastic dependencies in the eigenvalues, so that the behavior of both algorithms would be changed fundamentally. Thereby the presented bounds would be rendered invalid. Since the presented GLR algorithm is computationally expensive, it would be interesting to investigate efficient heuristics which can be used in practical scenarios. Additionally, it would be interesting to extend the simple channel model to more realistic fading channel models. Finally, it may be relevant to examine other test statistics based on the (eigenvalues of) the sample covariance matrix for quickest spectrum sensing.

%% file: figures/QD.pdf_tex
\begingroup%
  \makeatletter%
  \providecommand\color[2][]{%
    \errmessage{(Inkscape) Color is used for the text in Inkscape, but the package 'color.sty' is not loaded}%
    \renewcommand\color[2][]{}%
  }%
  \providecommand\transparent[1]{%
    \errmessage{(Inkscape) Transparency is used (non-zero) for the text in Inkscape, but the package 'transparent.sty' is not loaded}%
    \renewcommand\transparent[1]{}%
  }%
  \providecommand\rotatebox[2]{#2}%
  \ifx\svgwidth\undefined%
    \setlength{\unitlength}{236.1359375bp}%
    \ifx\svgscale\undefined%
      \relax%
    \else%
      \setlength{\unitlength}{\unitlength * \real{\svgscale}}%
    \fi%
  \else%
    \setlength{\unitlength}{\svgwidth}%
  \fi%
  \global\let\svgwidth\undefined%
  \global\let\svgscale\undefined%
  \makeatother%
  \begin{picture}(1,0.49398292)%
    \put(0,0){\includegraphics[width=\unitlength]{QD.pdf}}%
    \put(0.00115135,0.32577897){\color[rgb]{0,0,0}\makebox(0,0)[lb]{\smash{$\hypone$}}}%
    \put(0.00115135,0.09540319){\color[rgb]{0,0,0}\makebox(0,0)[lb]{\smash{$\hypzero$}}}%
    \put(0.84473324,0.02086985){\color[rgb]{0,0,0}\makebox(0,0)[lb]{\smash{$\timeidx$}}}%
    \put(0.45391717,0.02086985){\color[rgb]{0,0,0}\makebox(0,0)[lb]{\smash{$\tchange$}}}%
    \put(0.60419382,0.02086985){\color[rgb]{0,0,0}\makebox(0,0)[lb]{\smash{$\tdetect$}}}%
    \put(0.3817451,0.42721019){\color[rgb]{0,0,0}\makebox(0,0)[lb]{\smash{detection delay}}}%
  \end{picture}%
\endgroup%

%% file: figures/pdf_plot.tex
\begin{tikzpicture}
\begin{axis}[%
width=\figurewidth,
height=\figureheight,
xmin=1,
xmax=2.5,
ymin=0,
ymax=9,
ylabel={$\pdf_0(\pdfvar),~\pdf_1^\snr(\pdfvar)$},
xlabel=$\pdfvar$,
legend cell align=left,
legend entries={noise only, $\snr = -15~\dB$, $\snr = -12~\dB$, $\snr = -9~\dB$, $\snr = -6~\dB$, $\snr = -3~\dB$}]

\addplot [
color=black,
thick,
solid
]
table {figures/pdf_plot_data_0.dat};

\addplot [
color=red,
thick,
densely dashed
]
table {figures/pdf_plot_data_1.dat};

\addplot [
color=blue,
thick,
densely dotted
]
table {figures/pdf_plot_data_2.dat};

\addplot [
color=green,
thick,
densely dashdotted
]
table {figures/pdf_plot_data_3.dat};

\addplot [
color=cyan,
thick,
densely dashdotdotted]
table {figures/pdf_plot_data_4.dat};

\addplot [
color=orange,
thick,
loosely dashed]
table {figures/pdf_plot_data_5.dat};

\end{axis}
\end{tikzpicture}%

%% file: figures/pdf_verification_plot.tex
\begin{tikzpicture}
\begin{axis}[%
width=\figurewidth,
height=\figureheight,
xmin=1,
xmax=1.35,
ymin=0,
ymax=9,
ylabel={$\pdf_0(\pdfvar),~\pdf_1^\snr(\pdfvar)$},
xlabel=$\pdfvar$,
legend cell align=left,
legend entries={noise only, $\snr = -15~\dB$}
]

\addplot [
color=black,
thick,
solid
]
table {figures/pdf_verification_plot_data_0.dat};

\addplot [
color=red,
thick,
densely dashed
]
table {figures/pdf_verification_plot_data_1.dat};

\addplot [
color=black,
mark=o,
only marks
]
table {figures/pdf_verification_plot_data_emp_0.dat};

\addplot [
color=red,
mark=o,
only marks
]
table {figures/pdf_verification_plot_data_emp_1.dat};

\end{axis}
\end{tikzpicture}%

%% file: figures/roc_plot.tex
\begin{tikzpicture}
\begin{axis}[%
width=\figurewidth,
height=\figureheight,
xmin=0,
xmax=1,
ymin=0,
ymax=1,
ylabel=$\pd$,
xlabel=$\pfa$,
legend cell align=left,
legend style={
at={(0.78,0.04)},
anchor=south},
legend entries={$\snr = -15~\dB$, $\snr = -12~\dB$, $\snr = -9~\dB$}
]

\addplot [
color=red,
thick,
densely dashed
]
table {figures/roc_plot_data_1.dat};

\addplot [
color=blue,
thick,
densely dotted
]
table {figures/roc_plot_data_2.dat};

\addplot [
color=green,
thick,
densely dashdotted
]
table {figures/roc_plot_data_3.dat};

\addplot [
color=red,
mark=o,
only marks
]
table {figures/roc_plot_data_simu_1.dat};

\addplot [
color=blue,
mark=o,
only marks
]
table {figures/roc_plot_data_simu_2.dat};

\addplot [
color=green,
mark=o,
only marks
]
table {figures/roc_plot_data_simu_3.dat};

\end{axis}
\end{tikzpicture}%

%% file: figures/QD_t_d_plot.tex
\begin{tikzpicture}
\begin{axis}[%
width=\figurewidth,
height=\figureheight,
xlabel={$\threshold$},
ylabel={$\td$},
xmin=0,
xmax=10,
ymin=0,
legend style={at={(0.05,0.78)}, anchor=west},
legend cell align=left
]

\addplot [
color=red,
thick,
solid
]
table {figures/QD_plot_data_theo_t_d.dat};
\addlegendentry{bound};

\addplot [
color=black,
thick,
densely dashed
]
table {figures/QD_plot_data_cusum_t_d.dat};
\addlegendentry{CUSUM};

\addplot [
color=blue,
thick,
densely dotted
]
table {figures/QD_plot_data_glr_t_d.dat};
\addlegendentry{GLR};

\end{axis}
\end{tikzpicture}%

%% file: figures/QD_t_fa_plot.tex
\begin{tikzpicture}
\begin{semilogyaxis}[%
width=\figurewidth,
height=\figureheight,
xlabel={$\threshold$},
ylabel={$\tfa$},
xmin=0,
xmax=6,
ymin=1,
legend style={at={(0.08,0.78)}, anchor=west},
legend cell align=left
]

\addplot [
color=red,
thick,
solid
]
table {figures/QD_plot_data_theo_t_fa.dat};
\addlegendentry{bound};

\addplot [
color=black,
thick,
densely dashed
]
table {figures/QD_plot_data_cusum_t_fa.dat};
\addlegendentry{CUSUM};

\addplot [
color=blue,
thick,
densely dotted
]
table {figures/QD_plot_data_glr_t_fa.dat};
\addlegendentry{GLR};

\end{semilogyaxis}
\end{tikzpicture}%

%% file: figures/glr_initial_effects.tex
\begin{tikzpicture}
\begin{axis}[
width=\figurewidth,
height=\figureheight,
ylabel={$\cusum(\blockidx),~\glr(\blockidx)$},
xmin=0,
xmax=140,
ymin=0,
legend style={at={(0.795, 0.025)}, anchor=south},
legend cell align=left,
name={upper plot}
]

\addplot [
color=black,
thick,
solid
]
table {figures/glr_initial_effects_data_cs.dat};
\addlegendentry{CUSUM};

\addplot [
color=red,
thick,
densely dashed
]
table {figures/glr_initial_effects_data_gl.dat};
\addlegendentry{GLR};
\end{axis}

\begin{axis}[
at={(upper plot.below south west)},
yshift=-0.3cm,
anchor=north west,
width=\figurewidth,
height=\figureheight,
xmin=0,
xmax=140,
ymin=-20,
ymax=-8,
ylabel={$\glrsnr$},
separate axis lines,
axis y line*=left,
legend cell align=left
]

\addplot [
color=blue,
thick,
densely dotted
]
table {figures/glr_initial_effects_data_gl_snr.dat};
\label{pgfplots:gl_snr};
\end{axis}

\begin{axis}[%
at={(upper plot.below south west)},
yshift=-0.3cm,
anchor=north west,
width=\figurewidth,
height=\figureheight,
xmin=0,
xmax=140,
ymin=1,
ymax=10,
axis y line*=right,
xlabel={block index $\blockidx$},
ylabel={$\glrvar$}
]
\addlegendimage{color=blue,thick,densely dotted};
\addlegendentry{$\glrsnr$};

\addplot [
color=green,
thick,
densely dashdotted
]
table {figures/glr_initial_effects_data_gl_k.dat};
\addlegendentry{$\glrvar$};

\end{axis}
\end{tikzpicture}%

%% file: figures/glr_performance.tex
\newcommand{\BDpd}{0.9269}
\newcommand{\BDpfa}{0.0147}
\newcommand{\BDNs}{100000}

\begin{tikzpicture}
\begin{axis}[%
width=\figurewidth,
height=\figureheight,
xmin=0,
xmax=400000,
ymin=0,
ymax=1,
ylabel={$\pfa,~\pd$},
xlabel={GLR run-time},
legend cell align=left,
legend style={
at={(0.73,0.11)},
anchor=south},
legend entries={$\pfa$, {$\pd,\snr\!=\!-20\,\dB$}, {$\pd,\snr\!=\!-19\,\dB$}, {$\pd,\snr\!=\!-18\,\dB$}, {$\pd,\snr\!=\!-17\,\dB$}, {$\pd,\snr\!=\!-16\,\dB$}, {$\pd,\snr\!=\!-15\,\dB$}}
]

\addplot [
color=black,
thick,
solid
]
table {figures/glr_performance_p_fa_data.dat};

\addplot [
color=red,
thick,
densely dashed
]
table {figures/glr_performance_p_d_data_1.dat};

\addplot [
color=blue,
thick,
densely dotted
]
table {figures/glr_performance_p_d_data_2.dat};

\addplot [
color=green,
thick,
densely dashdotted
]
table {figures/glr_performance_p_d_data_3.dat};

\addplot [
color=cyan,
thick,
densely dashdotdotted
]
table {figures/glr_performance_p_d_data_4.dat};

\addplot [
color=orange,
thick,
loosely dashed
]
table {figures/glr_performance_p_d_data_5.dat};

\addplot [
color=teal,
thick,
loosely dotted
]
table {figures/glr_performance_p_d_data_6.dat};

%
%

\addplot [
color=gray,
mark=*,
mark size=1.5,
only marks
]
coordinates {
(\BDNs, \BDpd)
};

\addplot [
color=gray,
solid
]
coordinates {
(0, \BDpd)
(\pgfkeysvalueof{/pgfplots/xmax}, \BDpd)
};

\addplot [
color=gray,
mark=*,
mark size=1.5,
only marks
]
coordinates {
(\BDNs, \BDpfa)
};

\addplot [
color=gray,
solid
]
coordinates {
(0, \BDpfa)
(\pgfkeysvalueof{/pgfplots/xmax}, \BDpfa)
};

\draw[gray]
(axis cs:\BDNs,\pgfkeysvalueof{/pgfplots/ymin})
-- (axis cs:\BDNs,\pgfkeysvalueof{/pgfplots/ymax});

\end{axis}
\end{tikzpicture}

%% file: figures/glr_threshold_spread.tex
\newcommand{\GLRthreshA}{3.0}
\newcommand{\GLRthreshB}{4.5}
\newcommand{\GLRthreshC}{6.0}

\begin{tikzpicture}
\begin{axis}[%
width=\figurewidth,
height=\figureheight,
xmin=0,
xmax=200000,
ymin=0,
ymax=1,
ylabel={$\pfa,~\pd$},
xlabel={GLR run-time},
legend cell align=left,
legend style={
at={(0.77,0.465)},
anchor=south},
legend entries={{$\pfa,\threshold\!=\!\GLRthreshA$}, {$\pd,\,\threshold\!=\!\GLRthreshA$}, {$\pfa,\threshold\!=\!\GLRthreshB$}, {$\pd,\,\threshold\!=\!\GLRthreshB$}, {$\pfa,\threshold\!=\!\GLRthreshC$}, {$\pd,\,\threshold\!=\!\GLRthreshC$}}
]

\addplot [
color=black,
thick,
solid
]
table {figures/glr_threshold_spread_p_fa_data_1.dat};

\addplot [
color=red,
thick,
densely dashed
]
table {figures/glr_threshold_spread_p_d_data_1.dat};

\addplot [
color=blue,
thick,
densely dotted
]
table {figures/glr_threshold_spread_p_fa_data_2.dat};

\addplot [
color=green,
thick,
densely dashdotted
]
table {figures/glr_threshold_spread_p_d_data_2.dat};

\addplot [
color=cyan,
thick,
densely dashdotdotted
]
table {figures/glr_threshold_spread_p_fa_data_3.dat};

\addplot [
color=orange,
thick,
loosely dashed
]
table {figures/glr_threshold_spread_p_d_data_3.dat};

\end{axis}
\end{tikzpicture}